# Photonic Topological Insulator in Synthetic Dimensions

Eran Lustig[1†], Steffen Weimann[2†], Yonatan Plotnik[1], Yaakov Lumer[3], Miguel A. Bandres[1], Alexander Szameit[2] and Mordechai Segev[1*]

1. Physics Department and Solid State Institute, Technion-Israel Institute of Technology, Haifa 32000, Israel
2. Institut für Physik, Universität Rostock, Rostock 18059, Germany
3. Department of Electrical and Systems Engineering, University of Pennsylvania, Philadelphia, Pennsylvania 19104, USA

[†] These authors contributed equally to this work

*msegev@technion.ac.il

**Topological phases enable protected transport along the edges of materials, offering immunity against scattering from disorder and imperfections[1]. These phases were suggested and demonstrated not only for electronic systems, but also for electromagnetic waves[2–9], cold atoms[10–12], acoustics[13], and even mechanics[14]. Their potential applications range from spintronics and quantum computing to highly efficient lasers. Traditionally, the underlying model of these systems is a spatial lattice in two or three dimensions. However, it recently became clear that many lattice systems can exist also in synthetic dimensions which are not spatial but extend over a different degree of freedom[15,16]. Thus far, topological insulators in synthetic dimensions were demonstrated only in cold atoms[17–19], where synthetic dimensions have now become a useful tool for demonstrating a variety of lattice models that are not available in spatial lattices[20–24]. Subsequently, efforts have been directed towards realizing topological lattices with synthetic dimensions in photonics, where they are connected to physical phenomena in high-dimensions, interacting photons, and more[25–29]. Here we demonstrate experimentally the first photonic topological insulator in synthetic dimensions. The ability to study experimentally photonic systems in synthetic dimensions opens the door for a plethora of unexplored physical phenomena ranging from PT-symmetry[30,31], exceptional points[31–33] and unidirectional invisibility[34] to Anderson localization in high dimensions and high-**

**dimensional lattice solitons[16], topological insulator lasers in synthetic dimensions[35,36] and more. Our study here paves the way to these exciting phenomena, which are extremely hard (if not impossible) to observe in other physical systems.**

In the field of topological insulators, one of the most striking phenomena is the appearance of topological edge-states[1]. Topological edge-states are eigenstates of the system that reside in a topological band gap, are robust to disorder, immune to backscattering from defects and are localized at the edge of the lattice. In light of these properties, transport via topological edge-states is important both for its fundamental aspects and for potential technological applications. The concepts underlying topological edge-states extend much beyond condensed matter. In fact, topological edge-states were demonstrated in a variety of physical systems, ranging from electromagnetic waves both in the microwave regime[3,37] and at optical frequencies[8,9], to cold atoms[11,12], acoustics[13] and even mechanical systems[14]. Despite of their different manifestations, all of these topological insulators systems rely on spatial lattices, and the wavepackets occupying the lattice, whether they are electrons, photons or phonons, are subjected to gauge fields that give rise to the topological phenomena.

However, lattices may take other forms than a spatial arrangement of sites: they can be assigned to a ladder of atomic states, photonic cavity modes, or spin states[15,26,38]. Using one (or more) of these ladders in a non-spatial – synthetic – degree of freedom requires that the coupling between the synthetic sites and the resulting gauge fields are introduced to the system as an additional external perturbation. In contrast to the traditional topological insulators that are based on a spatial lattice, for topological insulators in synthetic dimensions the edge transport does not occur on the spatial edges of the system, but rather on the edges of the synthetic space. For example, the lowest and highest modes in a system can serve as synthetic edges. Based on this concept, a topological edge-state on a lattice with one spatial dimension and one synthetic dimension – the atomic spin state - was demonstrated in cold atoms[18,19]. In photonics, synthetic dimensions were proposed for generating solitons in higher dimensions [16], and were demonstrated in the context of topological pumping[39,40], where a quasi-periodic lattice was mapped onto a corresponding quantum Hall lattice with twice its spatial dimensions. However, such mapping does not allow edge transport that is immune to scattering and disorder in synthetic space: the hallmark of topological insulators. Theoretically, photonic topological insulators in synthetic dimensions were proposed in two pioneering works with the synthetic space realized through cavity modes[26,41], but thus far photonic topological insulators in synthetic dimensions have never been demonstrated in experiments.

The importance of photonic topological systems in synthetic dimensions relates to phenomena that otherwise are extremely hard (or impossible) to observe in other physical systems. Synthetic dimensions implemented specifically via modal space allow introducing arbitrary geometries and gauge fields that are not available in real space lattices. In real space lattices, the coupling between sites is induced by the spatial proximity of the sites to each other. This nearest-neighbor coupling severely limits the range of possible lattice geometries available. In contrast, the coupling between sites in synthetic space is induced externally, and applying the external perturbation corresponds to choosing the lattice coupling scheme and the gauge fields. This allows to produce lattices with unusual features such as long-range coupling, high dimensionality, interactions and other exotic phenomena,[20–24,26–29,40–43].

Here we experimentally demonstrate the first photonic topological insulator in synthetic dimensions. To do so, we introduce a new scheme to create a lattice with a synthetic dimension in the lattice-mode space. Our scheme supports the existence of topological states in a band-gap that extend over the bulk of a 2D lattice. To demonstrate such a photonic topological insulator in synthetic dimensions, we judiciously design a photonic lattice that realizes a topological insulator in synthetic dimensions endowed with topological edge-states, and directly observe the propagation of the topological protected edge state. These topological edge-states are not at the spatial edges of the system but at the edge of synthetic space, which occur in the bulk in real space. Finally, although we focus here on topological phenomena, our scheme to create synthetic lattices is modular and leads to new experimentally realizable systems that can exhibit other lattice phenomena in synthetic dimensions. The ability to study experimentally photonic systems in synthetic dimensions opens the door for various unexplored physical phenomena and applications, extending from PT-symmetry in synthetic dimensions, highly nonlocal nonlinearities emulating General Relativity, four-dimensional solitons, instabilities and chaos to topological insulator lasers[35,36] in synthetic dimensions, where the entire bulk exhibits topologically protected lasing.

We begin by describing our scheme of a two-dimensional waveguide array, engineered such that it has one synthetic dimension in modal space and one dimension in real space. Our study is performed on a photonic lattice, but the design is general and can be implemented in completely different physical systems such as cold-atoms, acoustics and more. We construct the photonic topological insulator in synthetic dimensions as follows. Consider first a one-dimensional array of $N$ evanescently-coupled waveguides arranged along the $y$ axis and propagating in the $z$ direction (Fig. 1a). This array has $N$ propagating eigenmodes, which propagate with different propagation constants. We engineer the coupling between the waveguides such that the propagation constants of the propagating modes are equally spaced (see Supplementary Information). Thus, they form an equally-spaced ladder of modes in a synthetic space, where in the lowest mode the waves in all waveguides are in-phase while at the highest mode the relative phase between each two adjacent waveguides is $\pi$. The eigenmodes of the lattice in Fig. 1.a are not coupled. For example, if a wavepacket occupying the first mode is launched at the input of the lattice, it will remain in the first mode for the entire propagation until reaching the output of the lattice (purple circle in Fig1.a). However, the transport in synthetic space requires coupling between modes. Thus, to couple the modes, we spatially oscillate the 1D waveguide array along its longitudinal dimension (z-dimension - Fig.1b). The period of the oscillations is designed to induce coupling between modes that have adjacent propagation constants, and creates a "lattice of modes" in the synthetic modal dimension. Here, for example, if a wavepacket occupying the first mode of the non-oscillating lattice is launched at the input of the oscillating lattice - it will couple to higher modes – will climb in the modal ladder and will be in a superposition of high modes at the output of the lattice (purple circle in Fig1.b)

Next, consider arranging M such oscillating one-dimensional arrays next to one another, in equal distances along the $x$ axis (Fig.2a-b). Here, the one-dimensional arrays (which are all oscillating in z) are now columns in the y-direction of a two-dimensional lattice (Fig. 2.b). This system can also be viewed as a two-dimensional lattice where one of its dimensions is a synthetic dimension. The first dimension is the ordinary spatial dimension $x$ (the horizontal axis in Fig. 2a-b), but the second dimension is the mode spectrum of each column (the vertical axis in Fig. 2a). For example, the site (3,5) in our synthetic lattice represents light occupying the third mode of the fifth real-space column in the horizontal direction x. We choose the placements of the waveguides in each column to correspond to a $Jx$ lattice, since it has the convenient property of equally spaced propagation constants for the modes[44,45]. Therefore we can couple every mode to the modes above and below it by driving the lattice at a single spatial frequency $\Omega$.

The lattice we just described already contains a synthetic dimension: the modal space. However, it does not yet display topological edge states. To do that, we now add a gauge field in the synthetic dimension, by introducing a phase shift in the oscillations between each two adjacent columns of oscillating waveguides (Fig. 2.c). That is, the oscillations in all columns are at the same frequency and amplitude, but the oscillations in adjacent columns differ by a constant nonzero phase. This phase difference manifests a magnetic field in the synthetic lattice, thereby opening a topological bandgap displaying topologically protected transport of edge states.

Our model is described in real space by the Hamiltonian:

$$H = -\sum_{m,n} t_{m,n}(z) c^\dagger_{m+1,n} c_{m,n} - \sum_{m,n} p_n c^\dagger_{m,n+1} c_{m,n} \exp(i d_{y,n} k_0 \Omega R \cos(\Omega z + \phi_m)) + H.C \quad (1)$$

where $c^\dagger_{m,n} c_{m,n}$ are creation and annihilation operators of the real space sites, $m = \{1,\dots,M\}, n = \{1,\dots,N\}$, $p_n$ and $t_{m,n}(z)$ are the coupling coefficients, $R$ is the amplitude of the oscillations in the longitudinal $z$ direction which plays the role of time, $k_0 = 2\pi n_0/\lambda$ is the wavenumber in the ambient medium, $d_{y,n}$ is the distance in $y$ between site $n$ and site $n+1$, $\Omega$ is the spatial frequency of the oscillations, $\phi_m$ is the phase of the oscillations of the $m$'th column, and $H.C$ stands for Hermitian conjugate. We emphasize that in this two-dimensional setting all columns oscillate parallel to the $y-z$ plane, at the same frequency $\Omega$ and the same amplitude $R$, but each column can oscillate at its own phase, $\phi_m$. It is the oscillation phase, $\phi_m$, that makes this lattice topological, by realizing an artificial gauge field in the synthetic lattice. The transformation to the lattice in synthetic space is carried out by the unitary matrix $U$ that diagonalizes the Hamiltonian $H(t_{mn}=0, R=0)$, that is, $U$ diagonalizes the Hamiltonian $H$ in Eq.1 when the coupling between columns and the radius of oscillation are set to be zero. Notice that $U$ converts the basis of representation to modes of separate non-oscillating columns. Using $U$, the Hamiltonian of the synthetic lattice $\widetilde{H}$ is obtained by: $\widetilde{H} = U^\dagger H U$, where here $H$ is the Hamiltonian with non-zero $R$ and $t_{mn}$.

Choosing $R, \phi_m$ and $\Omega$ corresponds to choosing the gauge fields in the Hamiltonian $\widetilde{H}$ (Supplementary Material). For our purposes, we choose $\phi_m = \phi m$, such that the oscillations have a phase shift $\phi$ between adjacent columns (Fig2.c). Consequently, light accumulates the phase difference $\phi$ upon encircling a plaquette in synthetic space (Fig2.a). Introducing a phase $\phi$ breaks $z$ reversal symmetry ($z$ plays the role of time in our system), and the chirality induced by the phase $\phi$ is an effective magnetic field in the synthetic space Hamiltonian $\widetilde{H}$ (see Supplemental Material). As a result, the spectrum of the synthetic lattice $\widetilde{H}$, and therefore of the real space lattice $H$, is that of a two-dimensional topological insulator. Figure 2.d displays the calculated eigenmodes of the Hamiltonian (Eq.1) when periodic boundary conditions are applied in the $x$ direction, and $R$, $\phi_m$ and $\Omega$ are chosen appropriately. The states marked in red are topological edge-modes in a topological bandgap, but they are not on the edge of the lattice in real space. Rather, they reside on the edge of the lattice in synthetic space (yellow frame in Fig. 2.a) and in the bulk of the lattice in real space (red glow in Fig. 2.b)

As explained earlier, the edges in synthetic space are not the edges of real space. The bottom edge of synthetic space, for example, means occupying the lowest mode of each column, and the lowest mode of each column is extended over the entire column in real space. Thus, a wavepacket occupying the bottom edge of synthetic space will actually be extended over the entire bulk of the lattice in real space, with the same phase at all lattice sites within each column. In the same vein, a wavepacket occupying the upper edge of the synthetic space is also extended over the entire lattice, but with $\pi$ phase difference between adjacent sites within each column. The left and right edges have a completely different nature than the bottom and upper edges. Here, occupying the left edge in synthetic space also means occupying the left edge in real space, since $x$ is the spatial dimension of our lattice, and likewise for the right edge of our lattice. Altogether, a wavepacket encircling the edges of the synthetic lattice will, for example, start its journey at the bottom edge of synthetic space (the lowest mode of each column), then propagates first on the lower edge of synthetic space - which covers the entire bulk in real space - all the way to the leftmost column, where it will stay at leftmost column but climb up the modal ladder starting from the first mode until reaching the highest mode. After reaching the highest mode, the wavepacket will start propagating rightwards over the entire bulk of real space but occupying the high lattice modes, until reaching the right edge. After that, the wavepacket will descend the modal ladder until it reaches the lowest mode, completing a full cycle around the edges of our synthetic lattice.

Just like the edge-states of a two-dimensional photonic Floquet topological insulator[8], the topological edge modes of our synthetic system are immune to disorder and reside in the gap. We verify this important feature of topological immunity both analytically and by simulating the propagation of beams in many disorder realizations (see Supplementary Material). Figure 2.d presents the Floquet spectrum (propagation constants vs transverse momentum) of the system without any disorder, and the topological edge-states are clearly seen in the gap. For comparison, Fig. 2.e shows the spectrum of the system with random disorder in the coupling between waveguides. The disorder here is strong - of the order of half the band gap. However, when the system is exposed to disorder, the edge-states may shift and deform, but they cannot vanish as long as the disorder is smaller than (approximately) half the band gap.

To study experimentally the evolution of the edge-states in our synthetic topological insulator, we propagate a paraxial beam at $\lambda$=633nm through the 2D lattice shown in Fig. 2b. we fabricated two lattices: one lattice with $\phi=\pi/4$ (topological) and one with $\phi=0$ (topologically trivial). The lattice consists of 7x11 waveguides fabricated in fused silica using the direct laser writing technique[46]. The light propagates according to the paraxial equation:

$$i\frac{\partial}{\partial z}\psi = -\frac{1}{2k_0}\nabla^2\psi(x,y,z) + \frac{k_0\Delta n(x,y,z)}{n_0}\psi(x,y,z) \qquad (2)$$

where $\psi(x,y,z)$ is the electric field envelope function defined by: $\boldsymbol{E}(x,y,z) = \psi(x,y,z)\exp(ik_0 z - i\omega t)\hat{x}$, $\boldsymbol{E}(x,y,z)$ is the electric field, $z$ is the propagation axis, the Laplacian $\nabla^2$ is restricted to the transverse $(x-y)$ plane; $k_0 = 2\pi n_0/\lambda$ is the wavenumber in the ambient medium; $\omega = 2\pi c/\lambda$ is the optical frequency; c and $\lambda$ are the velocity and wavelength of light respectively. Here, $-\Delta n(x, y, z)$ is the 'effective potential': the variation in the refractive index relative to the ambient refractive index of the medium, $n_0$. We launch a Gaussian beam into the lattice, excite the edge-state in synthetic space at $z = 0$, and measure the intensity at the output face of the lattice using a CCD camera.

Next, we explain the properties that are dictated by the topology of our synthetic lattice. In our system, for $\phi = \frac{\pi}{4}$ there are two counter-propagating edge-states in synthetic dimensions, one in the lowest bandgap, and the second in the upper bandgap (Fig2d-e). We focus on the state in the lowest bandgap since it has a lower spatial frequency and thus it is easier to excite with a Gaussian beam. The edge state we measure resides at some range of $k_x$ values, which may vary in location and size due to some inherent disorder in the fabrication (see for example edge state in Fig. 2.d compared to Fig. 2.e). Since $\phi$ is positive, this edge-state should propagate clockwise in synthetic space. Furthermore, in the lower bandgap there should be no edge-states propagating counter-clockwise in the synthetic space. Thus, a wavepacket exciting this synthetic-space edge-state should not scatter while passing the corner of synthetic space; rather, it should propagate along the leftmost column of the lattice in synthetic space. For comparison, we expect that the same wavepacket launched into the topologically-trivial system (with $\phi = 0$) to strongly scatter into the bulk of synthetic space.

We begin our measurements by coupling light to the edge-state in the topological lattice and comparing its evolution to that in the trivial lattice. A Gaussian beam is incident upon the input facet of the lattices, covering most of the sites of the columns at the center (Fig. 3.a,e). The beam is made oblique in the x-direction but it has uniform phase in the y-direction. The uniform phase in y makes the beam similar to the first mode in synthetic space, hence it mainly excites the lowest mode of each column, which corresponds to the bottom edge of the synthetic lattice (Fig. 3i). By varying the angle of incidence, we scan the $k_x$ space. We measure the intensity pattern at the output facet of the lattice after the beam has propagated $15 cm$ in the lattice. At angles where a topological edge-state does not exist in both lattices, we observe that the beam evolves into the bulk of synthetic space (see for example Fig. 3.b,f for $k_x = 0$). However, at a range of angles where a topological edge-state does exist in the topological lattice, we observe that the beam propagates to the left side of the lattice (see for example Fig. 3.c,d for $k_x \approx \frac{\pi}{10}$ and $\frac{\pi}{5}$), thus reaching the bottom left corner of synthetic space (Fig. 3.j). From this, we deduce that the topological state is within an angular range of $\Delta k_x = \frac{\pi}{5}$. For the same range of angles in the trivial case, light changes its location widely in the $x$ direction (Fig.3.g,h). In synthetic space, the light propagates clockwise on the bottom edge of synthetic space until it reaches the lowest mode of the leftmost column (purple circle in Fig. 3.i). This is confirmed by the intensity cross-section at the leftmost column (Fig. 3.j) and by interference measurements (Supplementary Material), compared to the calculated patterns. For comparison, at the same angles (Fig. 3.g,h) and in all other angles in the trivial lattice, the beam evolves into the bulk of synthetic space (blue arrow pointing up in Fig. 3.i) because the trivial lattice has no topological edge states. Since the beam is moving up in the modal dimension of the trivial lattice, it populates high lattice modes, as proved by the intensity cross sections (Fig 3.k).

After finding the topological edge state, we must examine that it indeed has its two prominent properties: that it is unidirectional, and that it does not scatter upon hitting the corner. To show this, we launch a narrow Gaussian beam with the width of roughly three lattice sites, at different regions of the lattice (Fig. 4a). We denote the center of the Gaussian beam with $x_0$, where $x_0 = 11$ means the beam is at the rightmost column, and $x_0 = 1$ is the Gaussian beam centered at the leftmost column (here, the rows in Fig.4 from top to bottom corresponds to $x_0 = 11, 8.5, 6, 3.5$ and 1, respectively). The narrow beam excites a narrower wavepacket on the synthetic space edge than previously (Fig. 4b). We chose the angle of the Gaussian beam to match the region in the Brillouin zone where the edge-state resides in the gap that was identified by scanning of the Brillouin zone described above. Then, we measure the output intensity pattern while moving the position of the Gaussian launch-beam across the $x$ axis, from the rightmost column ($x_0 = 11$) to the leftmost column ($x_0 = 1$). This corresponds to exciting the synthetic-space edge at different points along the propagation path of a beam evolving in the synthetic lattice (Fig. 4.b), and thus enables to study the dynamics in synthetic space.

The output beams for the different values of $x_0$ are presented in Fig. 4.c-d. Figure 4.c is the experimental image and Fig. 4.d displays simulation of a wavepacket propagating in our system on the edge of synthetic space. When the beam is launched at the right side of the lattice ($x_0 = 8.5 - 11$; rows: 1,2), we find that the output beam is spread and resides in the bulk. This corresponds to the beam propagating clockwise in synthetic space (Fig. 4.e; rows 1,2). When the beam is launched at the middle of the lattice ($x_0 = 6$; row 3), all the light accumulates in the leftmost column, mostly in the first mode (for further analysis of the experimental data see Supplemental Material). Upon moving the launch beam even more to the left ($x_0 = 1, 3.5$; rows: 4,5), higher modes are populated at the leftmost column of the lattice output (for further analysis of the experimental data see Supplemental Material). This finding demonstrates the unidirectionality of the edge-state: moving left causes the wavepacket to ascend the modal ladder, whereas moving right causes it to descend the ladder. As Fig.4 shows, the beam excites an edge state that moves only in one direction; that is, propagation of the edge state in the opposite direction does not occur both in the experiment and in simulations for any incident angle. At the same time, we observe that the beam does not scatter off the sharp (bottom-left) corner, which demonstrates topological protection. Altogether, these experiments prove that the propagation of the edge mode in our synthetic-space lattice possesses all the properties expected from topological edge modes, which in our lattice occur in synthetic space. We have therefore demonstrated a photonic topological insulator in synthetic space.

To conclude, we studied theoretically and experimentally a two-dimensional lattice that has one modal dimension and one spatial dimension, and that is subjected to a gauge field in synthetic space, creating a photonic topological insulator in synthetic space. We demonstrated that this topological lattice supports topological edge-states in the synthetic space. Our experimental platform serves as a building block for studying a variety of physical phenomena involving synthetic dimensions. For example, employing nonlinearities can lead to synthetic-space topological solitons, instabilities and chaos. Moreover, employing nonlocal nonlinearities (e.g., thermos-optical nonlinearities) links these ideas to synthetic-space General Relativity. Furthermore, the possibility of employing long-range coupling together with short-range coupling in our lattice by using several external perturbations at different frequencies leads to high dimensional effects[47] and to new unexplored lattice geometries. Finally, adding gain and loss to our synthetic space lattice would pave the way to non-Hermitian phenomena such as PT-symmetry, exceptional points, and lasing. This is just a handful of ideas that emerge from the experimental platform we have demonstrated here.

Funding: This work was supported by the German-Israeli DIP Program, and by an Advanced Grant from the ERC.

Authors Contributions: all authors contributed signifcantly to this work.

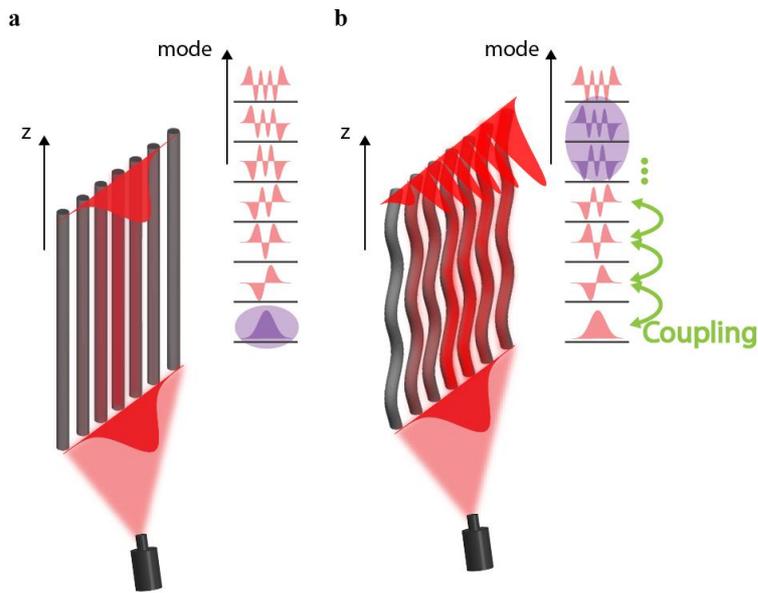

Figure1. **Forming a 1D lattice in a synthetic modal dimension.** (**a**) One-dimensional lattice with a spectrum of $N = 7$ eigenmodes with equally spaced propagation constants. (**b**) Oscillating the lattice in the longitudinal direction causes each eigenmodes of the straight lattice to couple to its nearest neighbors, forming a lattice of coupled modes.

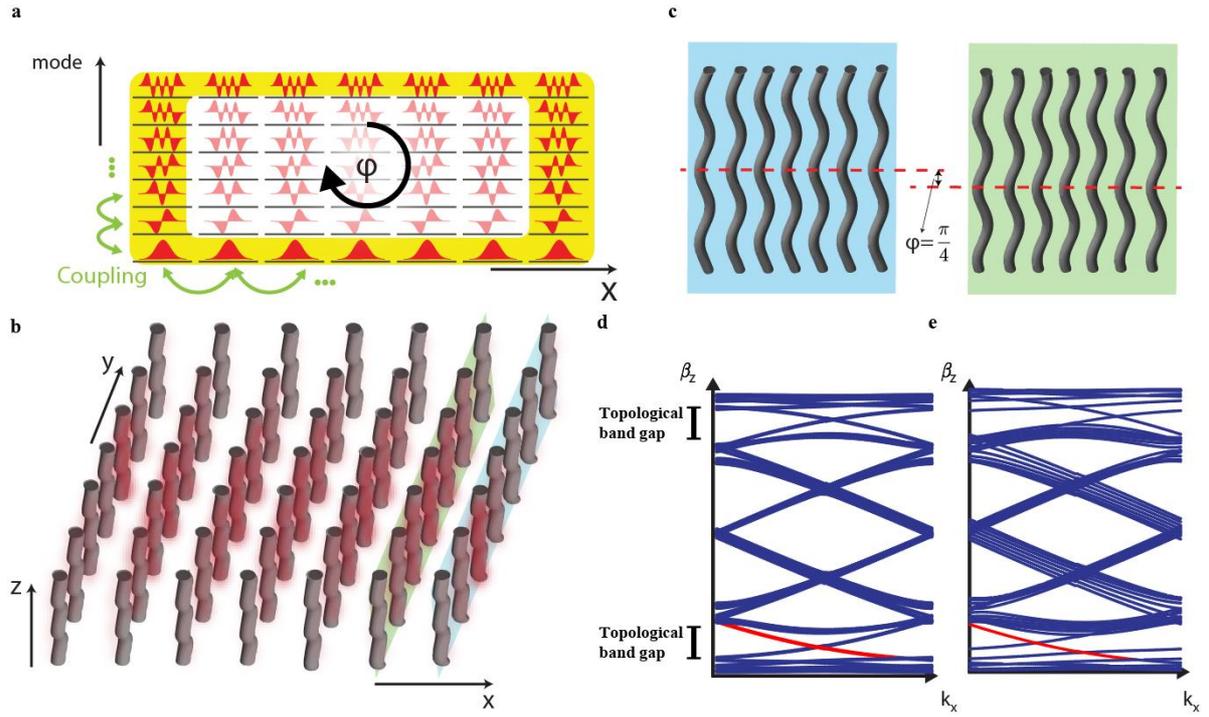

Figure 2. **The 2D synthetic space lattice. (a,b)** (a) is the synthetic space lattice corresponding to the two-dimensional lattice of waveguides presented in (b). The edge-state of the synthetic space (yellow in (a)) resides in the bulk of the waveguide array of (red in (b)). **(c)** The phase shift between each two adjacent columns of the waveguide array of (b). **(d)** The Floquet band structure of the lattice. The edge-state depicted in (a) and (b) is marked in red. **(e)** Floquet band structure as in (d) but with random disorder in the coupling between waveguides. The disorder mostly causes shifts and slight deformations in the dispersion curve of the edgestate, but does not close the topological gap, highlighting the immunity of the topological edge state to disorder.

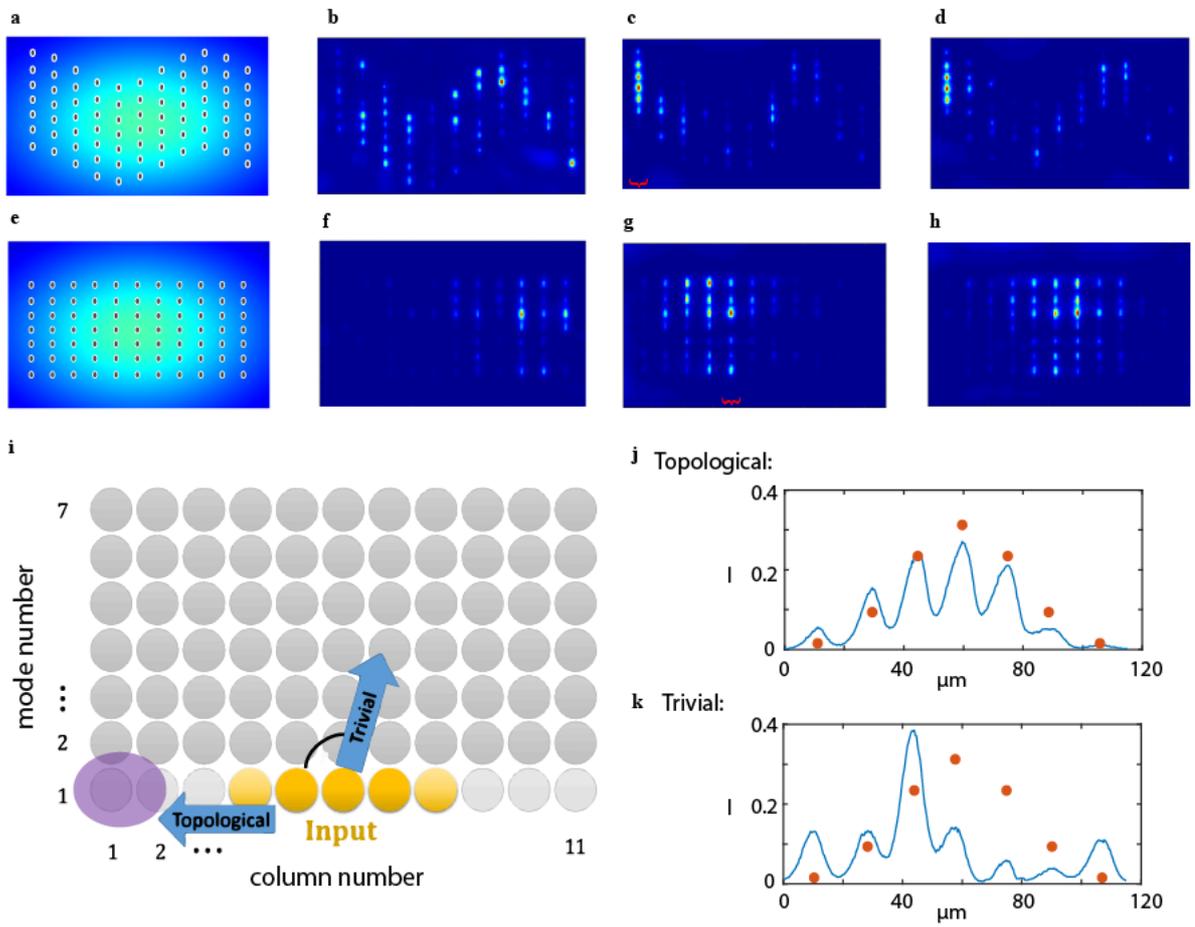

Figure 3. **Experimentally observed evolution of a topological edge-state in synthetic dimensions compared to the evolution in the topologically-trivial synthetic lattice.**

(**a-d**) Evolution in the topological lattice ($\phi = \frac{\pi}{4}$), for the Gaussian beam incident at the input facet of the topological lattice, as illustrated in (a). (**b**) Experimentally imaged output beam for the input of (a), at a launch angle that has no edge-states. (**c-d**) Experimentally imaged output beam for the lattice of (a) for angles where edge-states exist. (**e-h**) Evolution in the trivial lattice ($\phi = 0$), for the Gaussian beam incident at the input facet of the trivial lattice, as illustrated in (e). (**f-h**). Same angles as (b-d) respectively for the trivial lattice. (**i**) Illustrated evolution in synthetic space. The input beam (yellow) in (a-h) excites the lowest mode of the columns in the middle. In the topological lattice, when $k_x$ corresponds to an edge-state (images (c-d)), the beam evolves to the first mode of the leftmost column (purple). In the trivial lattice, the beam output is in the middle of the bulk but evolves to higher modes (**j**) Measured intensity cross-section (blue curve) at the red-marked columns in the topological (c) and trivial (g) lattices, compared with the calculated intensity profile of the first mode (orange dots).

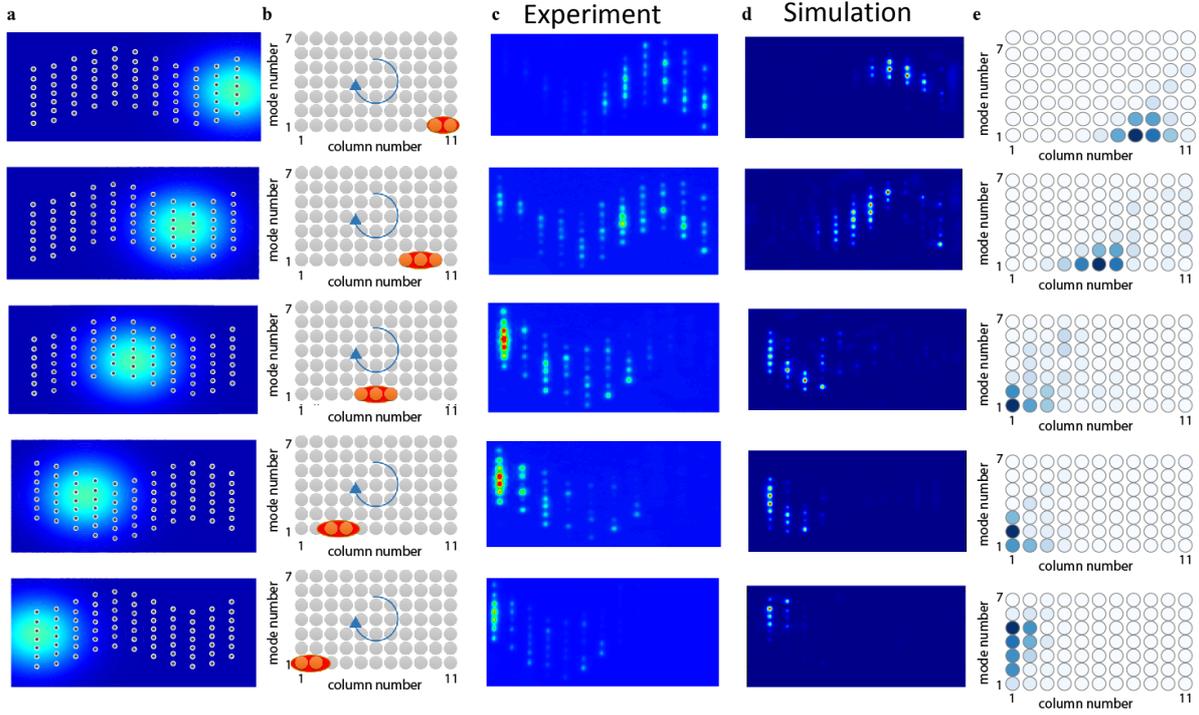

Figure 4: **Propagation of a wavepacket on the edge state of synthetic space.**
Each row correspond to a different point on the propagation path. **(a)** Illustration of the incident narrow Gaussian beam directed toward the input facet. **(b)** The corresponding input beam (red) in synthetic space. **(c,d)** Experimentally imaged and simulated beam intensity at the output, respectively. **(e)** Synthetic space description corresponding to the wavepacket propagation (d).

# Photonic Topological Insulator in Synthetic Dimensions
# Supplementary Information

Eran Lustig[1†], Steffen Weimann[2†], Yonatan Plotnik[1], Yaakov Lumer[3],

Miguel A. Bandres[1], Alexander Szameit[2] and Mordechai Segev[1*]

1. Physics Department and Solid State Institute, Technion-Israel Institute of Technology, Haifa 32000, Israel
2. Institut für Physik, Universität Rostock, Rostock 18059, Germany
3. Department of Electrical and Systems Engineering, University of Pennsylvania, Philadelphia, Pennsylvania 19104, USA

[†] *These authors contributed equally to this work*

## Supplementary notes 1

**Analytic derivation of the lattice in synthetic dimensions**

Here, we describe in details the interpretation of the waveguide array described in the main text as a lattice with one spatial dimension and one modal dimension. More specifically, we show that a two-dimensional waveguide array can be treated as a lattice in synthetic dimensions endowed with gauge fields, and that these gauge fields can give rise to edge-states in the synthetic space.

The equation governing our system is the paraxial equation for the propagation of light:

$$i\frac{\partial}{\partial z}\psi(x,y,z) = -\frac{1}{2k_0}\nabla^2\psi(x,y,z) + \frac{k_0\Delta n(x,y,z)}{n_0}\psi(x,y,z) \equiv H\psi(x,y,z) \quad (1)$$

where $\psi(x,y,z)$ is the envelope function of the $x$-component of the linealy polarized electric field defined by: $E(x,y,z) = \psi(x,y,z)\exp(ik_0z - i\omega t)$, $E(x,y,z)$ is the electric field, z is the propagation axis, the Laplacian, $\nabla^2$ is restricted to the transverse $(x - y)$ plane; with $n_0$ being the ambient refractive index of the medium, $k_0 = 2\pi n_0/\lambda$ is the wavenumber in the ambient medium; $\omega = 2\pi/\lambda$ is the optical frequency; $c$ and $\lambda$ are respectively the velocity and wavelength of light in vacuum and

$\Delta n(x, y, z)$ is the deviation from $n_0$ and $V(x, y, z) = \frac{k_0}{n_0}\Delta n(x, y, z)$ is the effective potential of Eq. (1).

We design $\Delta n(x, y, z)$ to form a two-dimensional lattice of waveguides, with each waveguide supporting one propagating mode. The structure of the lattice is:

$$V(x,y,z) = -\frac{k_0 \Delta n(x,y,z)}{n_0} = -\sum_{m,n} \frac{k_0 \Delta n_{site}(x - d_x m, y - d_{y,n} - R\sin(\Omega z + \phi_m), z)}{n_0} \quad (2)$$

where $\Delta n_{site}$ is the refractive index deviation (from the ambient refractive index of the medium) defining a single site, $d_{y,n}$ are the distances between the sites in the $y$-axis (columns), and $d_x$ are the distance between the sites in the $x$-axis (rows), $m = \{1, ..., M\}$ and $n = \{1, ..., N\}$ denote the waveguides of the spatial lattice. The waveguides oscillate in space only in the $y$ dimension, where $\Omega$ is the spatial frequency of the oscillations and $\phi_m$ is the phase of the oscillations of the $m$'th column (see Fig. 2(b) or Fig. 3(a) in the main text). For this derivation we treat our two-dimensional lattice (described by Eq.2) as being composed of one-dimensional waveguide arrays (columns) that are placed next to each other separated by a distance $d_x$. First we will analyze one of the one-dimensional waveguide arrays (columns) that compose the two-dimensional lattice, when it is isolated from the other columns ($d_x \to \infty$). One isolated column of the lattice is described by the effective potential and Hamiltonian:

$$V_{1D}(x,y,z) = -\sum_n \frac{k_0 \Delta n_{site}(x, y - d_{y,n} - R\sin(\Omega z + \phi_m), z)}{n_0} \quad (3)$$

$$H_{1D} = -\frac{1}{2k_0}\nabla^2 + V_{1D}(x,y,z) \quad (4)$$

To understand how the oscillations affect the propagation of the light within the waveguides, we change the frame of reference (change of coordinates) to the frame of the oscillating waveguides according to $x' = x, y' = y - R\sin(\Omega z + \phi_m), z' = z$. The propagation equation for the transformed wave function in the new coordinates $\psi'_{1D}(x', y', z')$ is:

$$i\frac{\partial}{\partial z}\psi'_{1D} = -\frac{1}{2k_0}\left(\vec{\nabla} + i\vec{A}(x', y', z')\right)^2 \psi'_{1D} - k_0\frac{\Omega^2}{2}R^2\cos^2(\Omega z' + \phi)\psi'_{1D} + \frac{k_0\Delta n(x', y', z')}{n_0}\psi'_{1D} \quad (5)$$

where $\vec{A} = k_0\Omega R\cos(\Omega z' + \phi_m)\hat{y}$. It is possible to eliminate one of the terms in Eq.(5) by using the gauge transformation: $\chi_{1D}(x', y', z') = e^{-i\int k_0\frac{\Omega^2}{2}R^2\cos^2(\Omega z + \phi_m)dz}\psi'_{1D}$:

$$i\frac{\partial}{\partial z}\chi_{1D} = -\frac{1}{2k_0}\left(\vec{\nabla} + i\vec{A}\right)^2 \chi_{1D} + \frac{k_0\Delta n}{n_0}\chi_{1D} \equiv H'_{1D}\chi_{1D} \quad (6)$$

The waveguides of our array are evanescently coupled to each other, and the propagating modes are bounded modes that are localized at the locations of the waveguides. Thus, a tight binding approximation can be employed, and together with Peierls substitution and nearest-neighbors (NN) approximation, we get the equation:

$$i\frac{\partial}{\partial z}c_n(z) = p_n e^{i\varphi_n}c_{n+1} + p_{n-1}e^{i\varphi_{n-1}}c_{n-1} \quad (7)$$

where $c_n$ are the tight binding coefficients of $\chi$, $\chi(x', y', z') = \sum_i c_i w_i(x', y', z')$, $w_i(x', y', z')$ are the basis functions of the tight binding approximation and $\varphi_n = \int_0^{d_{y,n}} \vec{A}(\vec{r})d\vec{r} = d_{y,n} k_0\Omega R\cos(\Omega z' + \phi_m)$. The coupling coefficient between the waveguides in each column is the parameter $p_n$ that corresponds to the coupling between the waveguide indexed $n$ and the waveguide indexed $n+1$. The coupling $p_n$ can be calculated directly from the distance between waveguides $d_{y,n}$ and knowledge on the shape of the waveguides. We set the couplings $p_n$ such that the propagating modes (eigenmodes) of the column we are analysing (in the absence of modulation, or $R = 0$) have equally spaced propagation coefficients in $k_z$, the propagation constant

(which serves as the spatial equivalent of energy). The coupling between waveguide $n$ and $n+1$ of the column is:

$$p_n = \frac{p}{2}\sqrt{j-n}\sqrt{j+n+1} \qquad (8)$$

where $p$ is a constant value with dimensions of spatial frequency, and $j = \frac{N+1}{2}$. This choice of couplings $p_n$ are known as the couplings of a $J_x$ lattice [references 53 and 54 in the main text].

Equation (7) endowed with the couplings in Eq.(8) describes a one-dimensional waveguide array (one column) that has nearest neighbour (NN) couplings. Next, we show that this waveguide array can be treated as a lattice with NN coupling in modal dimension, where the modes are those of a $J_x$ lattice. We Show this by changing the basis in which we write the Hamiltonian. The change in basis is the transformation between real space and synthetic space.

To describe the transformation between real space and synthetic space (a lattice of modes) consider the following unitary transformation:

$$\begin{pmatrix} c_1 \\ \vdots \\ c_n \\ \vdots \\ c_N \end{pmatrix} = \begin{pmatrix} a_1 \\ \vdots \\ a_n e^{i[\sum_{\alpha=1}^{n-1} d_{y,\alpha}]k_0 R\Omega \cos(\Omega z + \phi_m)} \\ \vdots \\ a_N e^{i[\sum_{\alpha=1}^{N-1} d_{y,\alpha}]k_0 R\Omega \cos(\Omega z + \phi_m)} \end{pmatrix} \equiv U \begin{pmatrix} a_1 \\ \vdots \\ a_n \\ \vdots \\ a_N \end{pmatrix} \qquad (9)$$

Where $U^\dagger U = I$, Applying the transformation of Eq.(9) on Eq.(7) and denoting $D_n = \sum_{\alpha=1}^{n-1} d_{y,\alpha}$ gives:

$$i\frac{\partial}{\partial z}a_n = D_n k_0 R\Omega^2 \cos(\Omega z + \phi_m)a_n + p_n a_{n+1} + p_{n-1} a_{n-1} \equiv [H_0 + H_1(z)]a_n \qquad (10)$$

Transformation (9) shifts the phases $\varphi$ from the off-diagonal elements in Eq.(7) to the main diagonal in Eq.(10). This allows a convenient decomposition of the Hamiltonian into a static Hamiltonian $H_0$ and a $z$-dependent Hamiltonian $H_1(z)$, the total Hamiltonian, which is: $H_{1D} = H_0 + H_1(z)$. We note that $H_0$ is the Hamiltonian of the column when it is not oscillating ($R \to 0$). The synthetic lattice is the eigenmodes of the non-oscillating Hamiltonian $H_0$. Thus, to write $H_{1D}$ in synthetic space, we use a unitary transformation from real space to the eigenmodes of $H_0$. We define $V$ as the unitary matrix that diagonalizes $H_0$:

$$V^\dagger H_0 V = E_0 \qquad (11)$$

where $E_0$ is a diagonal matrix. Transforming $H_{1D}$ to the basis $V$ gives the Hamiltonian $H_{1D}$ in synthetic space:

$$H_{1D}^{synthetic} = V^\dagger(H_{1D})V$$
$$= pn b_n^\dagger b_n + p_n D_n k_0 R\Omega^2 \cos(\Omega z + \phi) b_{n+1}^\dagger b_n + p_{n-1} D_{n-1} k_0 R\Omega^2 \cos(\Omega z + \phi) b_{n-1}^\dagger b_n \quad (12)$$

Where $b_n$ is the occupancy of the mode $n$ of $H_0$. At this point, we go back to the description of the 2D system. We assume that $N$ columns (each described by $H_{1D}$ and $H_{1D}^{synthetic}$) are now placed next to each other at a distance $d_x$ to form a 2D lattice. The Hamiltonian of the 2D lattice with one synthetic dimension (the modes of the columns when they do not oscillate) and one spatial dimension (the column location on the spatial horizontal axis) is:

$$H_{2D}^{synthetic}$$
$$= pn b_{m,n}^\dagger b_{m,n} + p_n D_n kR\Omega^2 \cos(\Omega z + \phi_m) b_{m,n+1}^\dagger b_{m,n}$$
$$+ p_{n-1} D_{n-1} kR\Omega^2 \cos(\Omega z + \phi_m) b_{m,n-1}^\dagger b_{m,n} + t_{m,n}(z) b_{m+1,n}^\dagger b_{m,n}$$
$$+ t_{m-1,n}(z) b_{m-1,n}^\dagger b_{m,n} \qquad (13)$$

where the index $m$ is the column location (spatially). $H_{2D}^{synthetic}$ is the Hamiltonian in synthetic space since the index $m$ is for the spatial dimension, but the index $n$ is for the mode dimension. $H_{2D}^{synthetic}$ is a unitary transformation of the Hamiltonian of the 2D lattice in real space $H$, hence, the eigenvalues are the same for both Hamiltonians. By choosing $\phi_m = \phi m$ we obtain edge-states in synthetic dimensions. Other choices of $\Omega$ and $\phi_m$ will result in other gauge fields in synthetic dimensions. We chose $\Omega = p$, thus since the Hamiltonian is periodic in $\Omega$, the first term of Eq.(13) can be gauged out.

Next, we show why choosing $\phi_m = \phi m$ gives rise to edge-states in synthetic dimensions. Consider applying the unitary transformation: $u_{m,n} = b_{m,n} e^{i\Omega n z}$ on Eq.(13). After using the fast-rotating wave approximation, neglecting the terms with $e^{i\Omega z}$, we obtain the following Hamiltonian:

$$\widetilde{H}_{2D}^{synthetic} = \frac{1}{2} p_n D_n k_0 R\Omega^2 u_{m,n+1}^\dagger u_{m,n} e^{i\phi m} + \frac{1}{2} p_{n-1} D_{n-1} k_0 R\Omega^2 u_{m,n-1}^\dagger u_{m,n} e^{-i\phi m}$$
$$+ t_{m,n}(z) u_{m+1,n}^\dagger u_{m,n} + t_{m-1,n}(z) u_{m-1,n}^\dagger u_{m,n} \quad (14)$$

Hamiltonian (14) is a lattice with magnetic flux $\phi$ per plaquet. Here, the coupling coefficient are not constant since $p_n D_n$ and $t_{m,n}(z)$ are not constant in $z$. Importantly, had the coupling coefficients been constant, this model would have been identical to the Hofstadter model. As stated, $p_n D_n$ and $t_{m,n}(z)$ do vary with z, but they do not vary wildly. Specifically, if $d_x \gg R$, then $t_{m,n}(z) \approx$ const. Furthermore $p_n$ decreases when $D_n$ increases and vice-versa. Thus, although $p_n D_n$ is not constant in z, it does not vary strongly. Overall, for most experimentally feasible parameters Eq.(14) approximates to the Hofstadter model, and thus, have at least one topological edge-state, and the robustness against disorder displayed by the edge-states of the Hofstadter model. We verified the robustness of the topological edge states of our model by calculating the

Floquet band-structure of Eq.(14) with disorder in the coupling coefficients. Figures 2(d-e) show that, although the edge-states may shift in $k_x$, they do not disappear We confirm this result with many numerical simulations on the propagation in the presence of on-site disorder and disorder in the coupling coefficients (more on the disorder analysis in supplementary notes 3).

**Supplementary notes 2**

**Analysis of the experimentally measured dynamics of the wavepacket near and on the corner of synthetic space.**

In the main text we describe the propagation of a wavepacket on the edge of synthetic space. Here, we elaborate on the experimentally measured dynamics of the wavepacket near and on the corner of synthetic space. We specifically study the vicinity of the corner in synthetic space because it is the most complicated part of the dynamics. This section gives further results and conclusions related to the propagating wavepacket of Fig 4 in the main text. The propagation of the wavepacket on the edge of synthetic space, as described in the main text, starts at the lower edge of synthetic space (Fig 1.(a-b)), and continues clockwise until reaching the first mode of leftmost-column, which is the bottom-left corner of synthetic space. At this point of the propagation, we expect that the leftmost column will be occupied by only the first mode. To see this, we measure the intensity distribution of the light exiting the lattice in the left-most column, and compare it both in intensity (Fig.1c) and phase (Fig1.d) to the theoretically calculated first mode. The results match with high precision.

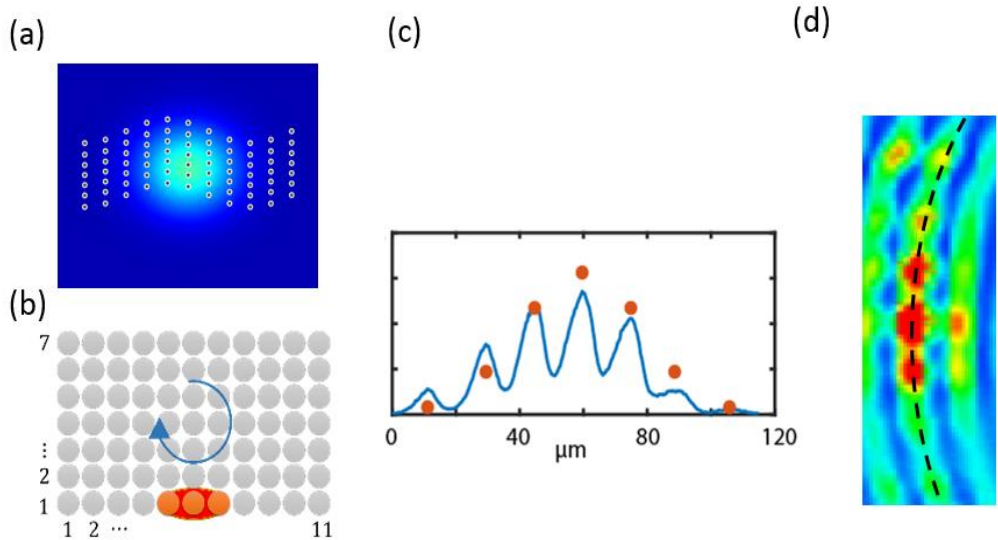

Figure 1: **Light occupancy in the leftmost column when a Gaussian beam is launched from the middle of the lattice**. (a-b) The initial Gaussian beam as launched in both real space and in synthetic space. (c) Measured intensity cross-section (blue) at the left-most column of the lattice output compared with the calculated intensity profile of the first mode (range dots). (d) Measured interference of the lattice output at the leftmost column with the input beam showing that the 7 peaks are all in phase (the dashed line is in equal phase in the input beam), indicating that the light at the leftmost column occupies the lowest spatial mode of the column.

After reaching the corner, we expect of the wavepacket to continue its propagation by staying at the leftmost column (instead of coupling to other columns as in the trivial case) and start occupying higher modes in the leftmost column. In Fig.2 we show the measured output light for the case in which the initial Gaussian beam is launched at the left side of the lattice, representing a later stage in the wavepacket's propagation (Fig2.a,b)

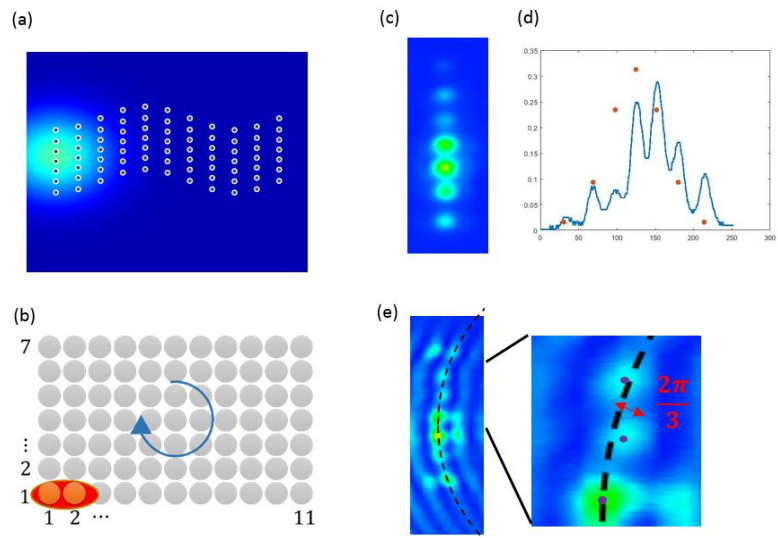

Figure 2: **Light occupancy in the leftmost column when Gaussian beam is launched from the left side of the lattice** (a-b) The initial Gaussian beam, as launched in both real space and in synthetic space. (c) Measured intensity distribution at the leftmost column. (d) Cross-section (blue) of the intensity pattern at the left-most column of the lattice output compared with the calculated intensity profile of the first mode. (d) Measured interference pattern at the lattice output at the leftmost column, showing that, unlike for Fig 1d above, the 7 intensity peaks are not in-phase, and therefore occupy higher modes of the array.

We measure the intensity (Fig2.c,d) and the phase (Fig2.e) of the light occupying the leftmost column. We observe that the intensity does not match the first mode, and that there is a phase difference of up to $\sim \frac{2\pi}{3}$ between adjacent sites, meaning that modes higher than the first are occupied at this point of evolution in the lattice, as expected from the simulations. By estimating the mode occupancy of Fig2.c according to phase and amplitude more than 50 percent of the light is in higher modes.

**Supplementary notes 3**

**Numerical analysis of the theoretical model**

In the main text, we present a new scheme for achieving a topological insulator in synthetic dimensions using a two-dimensional oscillating lattice. Here we give further details on this scheme. As mentioned in the main text, we can describe our model with the tight-binding approach. The tight binding equation for our scheme is Eq.(1) from the main text. In supplementary notes 1, we explain why Eq.(1) from the main text describes our lattice of waveguides, and how it relates to topological phenomena in synthetic dimensions. Here, we give further details on this model by studying it numerically.

Before we start, we will first describe how to choose the parameters of Eq.(1) from the main text such that it will fit the desired model in synthetic dimensions. Equation 1 of the main text can represent many models in synthetic dimensions upon different choices of $R$, $\Omega$, $p_n$ and $\phi_m$.

Choosing the couplings $p_n$ and the frequency $\Omega$:

The frequency $\Omega$ should correspond to the spacing between the different modes in the propagation constant $k_z$ of each column (which plays the role of energy in the analogous quantum problem). If the couplings $p_n$ are chosen according to Eq.(8) of supplementary notes 1, then the modes are equally spaced in $k_z$, and $\Omega$ is chosen according to that spacing. In the case of constructing a more elaborated lattice or a lattice that combines both long range couplings and NN couplings, one can oscillate the sites at a superposition of several frequencies $\{\Omega_i\}$.

Choosing the amplitude of oscillations $R$:

The amplitude of the oscillations $R$ corresponds to the strength of the coupling between the modes in synthetic space. Here, we choose $R$ such that it is large enough for having significant evolution dynamics in the finite propagation length in our experiments, and small enough such that it will not induce high order effects that obstruct the dynamics.

Choosing the phase $\phi_m$ :

The phase $\phi_m$ is the phase per plaquet in synthetic space.

After choosing the parameters that match the gauge fields corresponding to the Hofstadter model, one can study the model both as an eigenvalue equation and obtain the spectrum of the Hamiltonian, and as an evolution equation, and simulate the propagation of a wavepacket in this synthetic lattice. We carry out both of these studies.

We start by solving the Floquet eigenvalues and eigenvectors of Eq.(1) from the main text for a finite 20 by 20 lattice with $\phi = \pi/4$ (in the main text the spectrum is for a lattice of the same size as the experimental lattice, which is 7 sites in the columns and periodic boundary condition with $\phi = \pi/8$ ). The eigenvalues and one eigenvector are shown in Fig. 3. The spectrum clearly shows the existence of the edge-states, which reside in the bulk of real space, but on the edge of synthetic space.

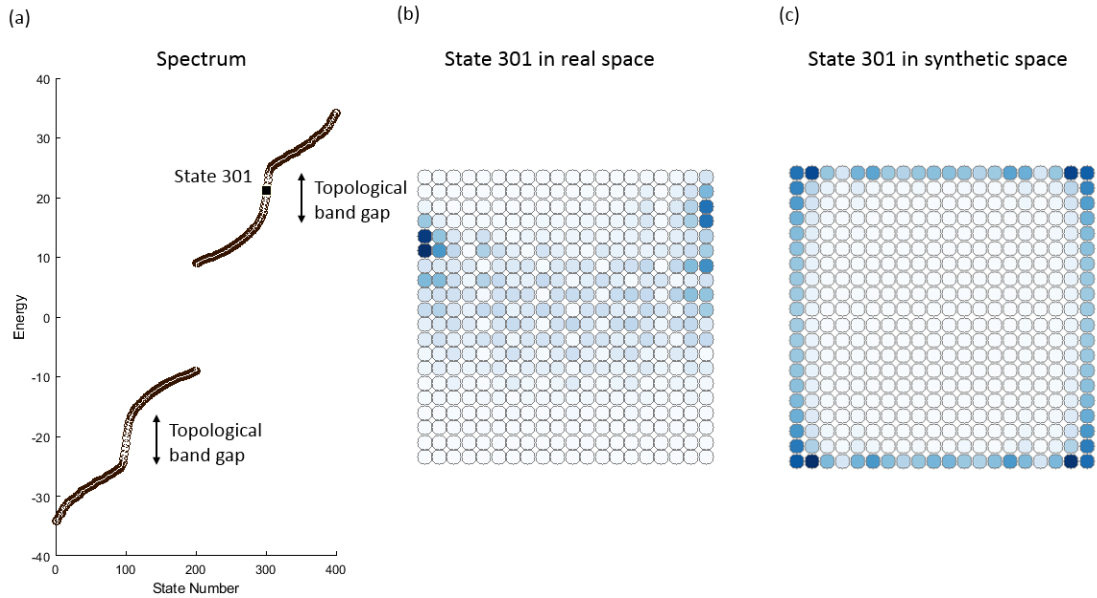

Figure 3: **Floquet spectrum of Equation 1 from the main text.** (a) The eigenvalues obtained numerically displaying two topological gaps. (b) Real space amplitude distribution of an eigenstate residing in the gap (state 301 marked in (a)). (c) Same eigenstate as in (b), as it appears in synthetic space (unitary transformation of (b)) show clearly that state 301 is an edge-state).

As mentioned, it is possible to simulate the propagation of a pulse propagating on the edge state. Figure 4 shows the propagation of a wavepacket in 3 different propagation planes. The middle image in the upper row in Fig.4 shows how the topological state resides in the bulk of the lattice. By designing larger lattices, it is possible to gradually change each column, and obtain topological edge-states in different shapes in the bulk. As an example, Fig.4(b) displays the real-space amplitude distribution of a topological edge-state in synthetic dimensions, as it is deformed by changing the couplings of each column

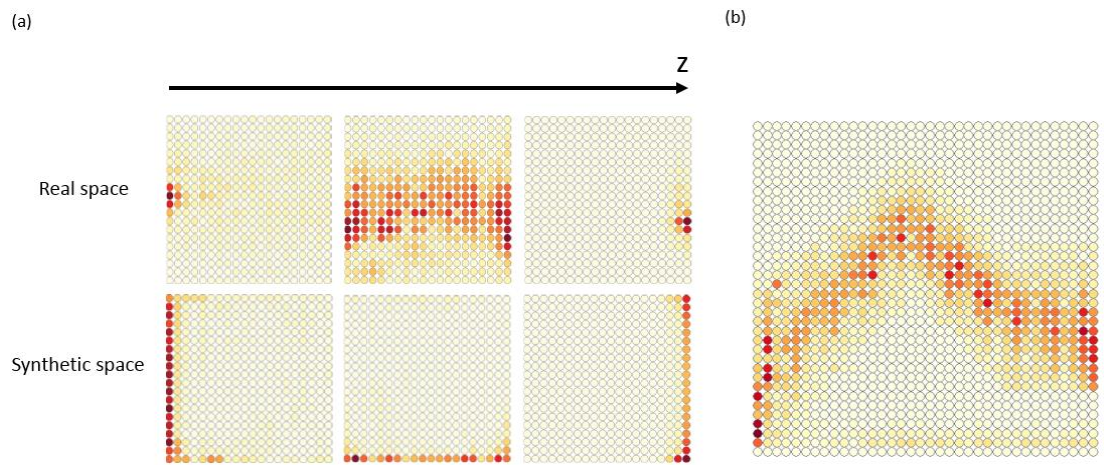

Figure 4: (a) The propagation of a wavepacket in synthetic dimensions at 3 different instances. The upper 3 images display the field distribution in real space, and the bottom 3 images gives the same distribution in synthetic space. (b) The amplitude of a deformed topological state obtained by changing the couplings between sites.

Finally, in supplemental notes 1, we showed that the bulk propagation in synthetic dimensions is mathematically equivalent to the Hofstadter model under certain approximations. Inevitably, it shares the same properties of the Hofstadter model but in synthetic dimensions. To that end, we simulate the lattice under random disorder in the couplings and in the on-site potential, with hundreds of realizations of the disorder, and as expected, the edge-states do not disappear under the influence of random noise. Furthermore, to make sure that random disorder in real space indeed behave the same in synthetic space, we calculate the topological invariant of the model under disorder - Bott index, and find that the Bott index remains an integer number even under disorder that is of the order of half the band gap.

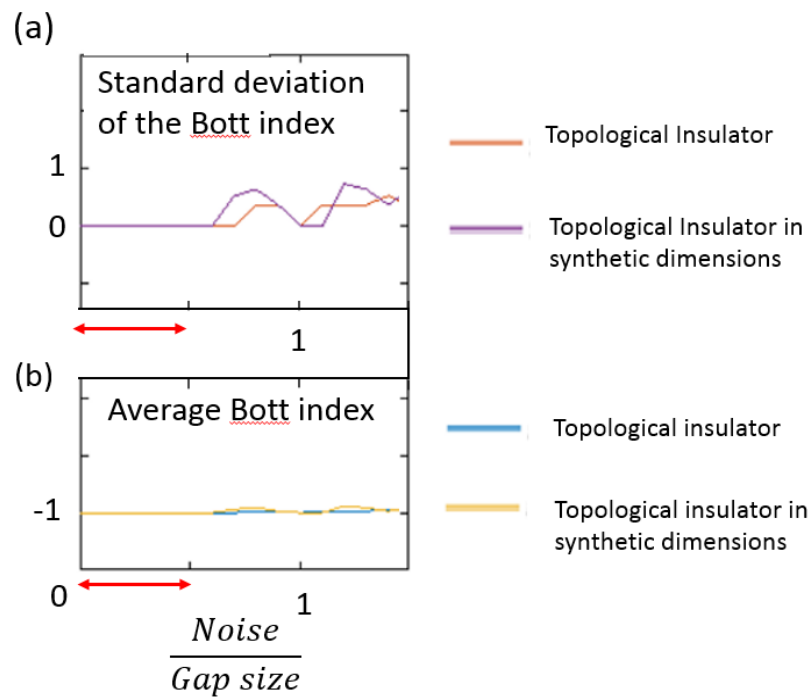

Figure 5: Bott index calculations of the model in synthetic dimensions compared to the Hofstadter model under random noise. **(a)** The standard deviation of the Bott index over 10 random disorder realizations both for the regular topological insulator, and for the topological insulator in synthetic dimensions described previously. The red arrow corresponds to half to band gap. **(b)** The average value of the Bott index for the same realizations as in (a).

Regarding disorder that is not random but particular (example: removing a single site), the robustness is not the same as for the regular topological insulator and for the topological insulator in synthetic dimensions. Since the unitary transformation between real space and synthetic space (Eq.11 in supplementary notes 1) is non-local, local disorder in real space becomes non-local in synthetic space, and vice-versa. Thus, although strong local defects may harm the robustness of the topological edge-state in synthetic dimensions, it is nevertheless robust to a range of non-local defects that harm regular topological edge-states but not the edge states of our topological synthetic lattice.

**Supplementary methods**

**Designing the experimental lattice in synthetic dimensions**

Here, we describe the process of designing the experimental photonic lattice we used in the main text. The process included several steps: First, we fabricated a single elliptical waveguide, and measured the intensity shape of the propagating modes in both its major axis and its minor axis (Fig.6.a,b). This process was done shortly prior to the fabrication of the lattice to avoid inaccuracies due to drifting of the system. Following this, we measured the coupling of two fabricated waveguides as a function of the distance between them and obtain exponential curves (Fig.6.c,d). The accuracy of the vertical coupling is more important than the horizontal one, since the vertical coupling is that of the $J_x$ lattice.

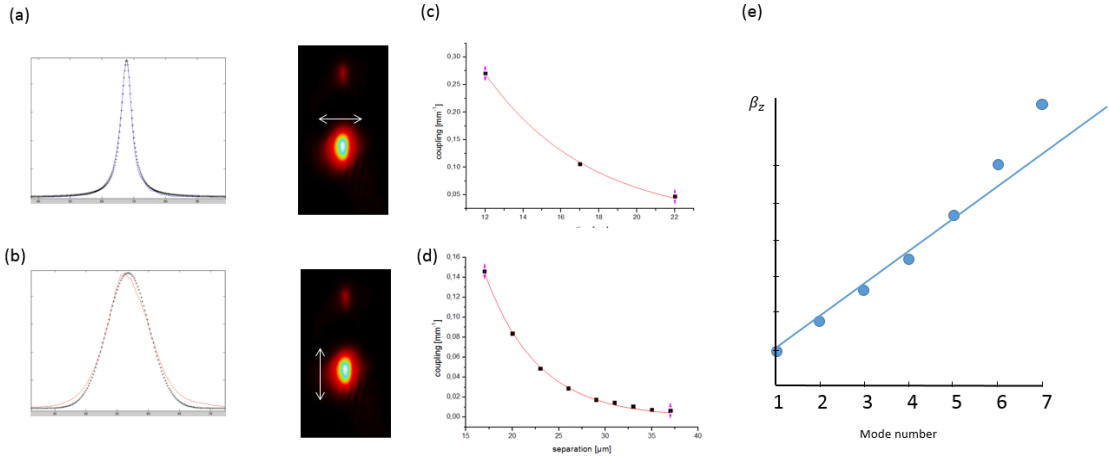

Figure 6: Stages in preparing the experimental model. (a) Horizontal cross section of the intensity of a propagating mode of a single waveguide. (b) Vertical cross section of the intensity of a propagating mode of a single waveguide. (c) Coupling vs distance in the horizontal direction. (d) Coupling vs distance in the vertical direction. (e) Modes of the $J_x$ lattice with the based on measured parameters (blue points).

Based on the measurements in Fig.6(a-d) we deduced the spatial distribution of the index of refraction of a single site $\Delta n_{site}$. Using $\Delta n_{site}$ we simulated a single $J_x$ lattice, the eigenvalues of the lattice are in Fig.6e. The first 5 eigenvalues are perfectly aligned, as required, but the modes 6 and 7 are tilting upward from the straight blue line. The reason for the deviation of modes 6 and 7 from the straight line is the relatively close proximity of the waveguides in the experiment which slightly violates the tight-binding approximation and therefore reduces the efficiency of the coupling of modes 6 and 7. Reducing the efficiency of the coupling in the 2 highest modes was a compromise done due to short propagation length in the experiment - separating the waveguides from each-other further reduces the average coupling and slows down the overall dynamics in $z$. Next, we used Beam Propagation Method to verify that the eigenmodes of a single oscillating $J_x$ lattice are all coupled to one another, and verified the existence of topological edge-state in synthetic dimension in the exact lattice we fabricated (these are the simulations in Fig.4.d in the main text). Finally, after simulating the system using data extracted from the experimental system, we fabricated the lattice and preformed the experiments described in the main text.